\documentclass[final,5p,times,twocolumn,authoryear]{elsarticle}

\usepackage{amssymb}
\usepackage{xcolor}
\usepackage{amsmath}
\usepackage[utf8]{inputenc}
\usepackage[T1]{fontenc}
\usepackage{tabularx}
\usepackage[pdfencoding=auto, psdextra]{hyperref}
\usepackage{comment}
\usepackage{url}
\usepackage{hyperref}
\usepackage{natbib} 

\journal{Physics of the Dark Universe }

\begin{document}

\begin{frontmatter}

\title{Testing Scale-Dependent Suppression of Structure Growth in the Linear Regime}

\author[label1]{Fernanda Oliveira}
\affiliation[label1]{organization={Observatório Nacional}, 
addressline={Rua General José Cristino, 77, São Cristóvão},
city={Rio de Janeiro},
postcode={20921-400},
state={RJ},
country={Brazil}}

\author[label4,label5]{Miguel A. Sabogal}
\affiliation[label4]{organization={Department of Physics, University of Trento},
addressline={Via Sommarive 14}, postcode={38123}, city={Povo (TN)},
country={Italy}}
\affiliation[label5]{organization={Trento Institute for Fundamental Physics and Applications (TIFPA)-INFN},
addressline={Via Sommarive 14}, postcode={38123}, city={Povo (TN)},
country={Italy}}

\author[label1]{Felipe Avila}

\author[label2,label3]{Rafael C. Nunes}
\affiliation[label2]{organization={Instituto de Física, Universidade Federal do Rio Grande do Sul}, 
addressline={},
city={Porto Alegre},
postcode={91501-97},
state={RS},
country={Brazil}}
\affiliation[label3]{organization={Divisão de Astrofísica, Instituto Nacional de Pesquisas Espaciais}, 
addressline={Avenida dos Astronautas 1758},
city={São José dos Campos},
postcode={12227-010},
state={SP},
country={Brazil}}

\author[label1]{Armando Bernui}

\begin{abstract}
We investigate recent reports of a suppression in the growth rate of cosmic structures inferred from analyses of the $[f\sigma_8](z)$ dataset. To address this issue, we explore the hypothesis that the evolution of matter clustering is more accurately described within the framework of scale-dependent modified gravity. We perform a joint analysis of $[f\sigma_8](z)$, cosmic chronometer $H(z)$ measurements, luminosity distance data, and CMB observations using Markov Chain Monte Carlo techniques to constrain the parameters of a scale-dependent cosmological model and investigate its impact on the evolution of $[f\sigma_8](z)$. 
Our results indicate that the suppression of the growth rate of large-scale structures is more pronounced during the matter-dominated era than in the dark-energy-dominated epoch. We find evidence for scale-dependent growth at a statistical significance of $2.2\, \sigma$. In addition, we constrain the $S_8$ parameter and find it to be consistent with the value inferred from the CMB observations of the Planck Collaboration. Overall, our analysis shows that $k$-dependent growth models provide a viable explanation for the observed clustering of matter without exacerbating the current cosmological tensions.
\end{abstract}

\begin{keyword}
Cosmology \sep Large-scale structure \sep Modified gravity

\end{keyword}

\end{frontmatter}

\section{Introduction}
\label{introduction}

The $\Lambda$CDM model has successfully reproduced a series of cosmological observations, such as luminosity distance data from Type Ia Supernovae (SN)~\citep{Riess1998, Perlmutter1998}, the angular power spectrum of the Cosmic Microwave Background (CMB) temperature fluctuations~\citep{Planck2018}, distance measurements of the Baryon Acoustic Oscillations (BAO)~\citep{DESI2}, and matter clustering~\citep{BOSS2016}.  
This model combines general relativity (GR) and the cosmological principle~\citep{Planck-isotropy, Perivolaropoulos2023, Jung2024, Franco2025c, Wu2025, Sanyal2026}, where almost 70\% of the universe composition being explained by dark energy, an unknown component responsible for the recent phase of 
accelerated expansion of the universe. 
The nature of dark energy is, along with other concerns, one of the open problems of the $\Lambda$CDM model~\citep{Weinberg2013, Perivolaropoulos2022, Dinda2025, DiValentino2025}. 

Several alternative cosmological models have been proposed and analyzed in recent years with the aim of addressing these problems~\citep{Chevallier2000,Linder2002, Clifton2011, BeltranJimenez2019, Akarsu2025, Avsajanishvili2026}, but a delicate issue concerning the growth rate of cosmic structures is still under debate~\citep{Peebles2022, Ruiz2015, Nguyen2023, Yang2025, Escobal2026}.In this work, we address a subject that has been little discussed in the literature; that is, we consider that the growth of cosmic structures depends on the scale $k$.In fact, in general, the cosmic growth rate function is obtained from the linear matter density contrast, $\delta_m(\textbf{r},a)$, a function that is scale-independent in the $\Lambda$CDM framework.
In this sense, the data from the growth rate of cosmic structures provide an important tool to test the potential scale dependence in alternative cosmological models.
Because many of these models have similar expansion histories, possible degeneracies among them are broken by combining data from the background and from the perturbed universe, that is, by resorting to data from the growth of cosmic structures, such as $f$ or $f\sigma_8$, to describe the matter clustering in the universe and to test the viability of alternative cosmological models~\citep{Oliveira2025b, Oliveira2025a, Wei2008, Knox2005, Basilakos2017, Akarsu2025}.

For small scales, i.e., for scales much smaller than the Hubble scale, the function $\delta_m(\textbf{r},a)$ can be obtained from a simple evolution equation called the growth equation, which is scale-independent (see, e.g.,~\cite{Nesseris2017, Perenon2019, Ribeiro23, Oliveira2024}).
However, at scales larger than 50 Mpc $h^{-1}$, it is appropriate to introduce a scale-dependent function in the growth equation~\citep{Dent2009, Perivolaropoulos2010}. In this regime, the solution of the scale-independent growth equation fails to reproduce the so-called GR solution~\citep{Dent2009}, making it necessary to use a scale-dependent function, $\xi(a, k)$, to approximate the solution of the growth equation to the GR solution~\citep{Dent2009, Nguyen2023}.

Our aim in this work is to test this approximation using cosmological data such as cosmic chronometers (CC), SN, the normalized growth rate $[f\sigma_8](z)$, and CMB through the Markov Chain Monte Carlo (MCMC) a suitable statistical method to obtain the best-fit values for cosmological parameters~\citep{Padilla2019,Gilks1995,Gelman2013, Bessa2021, DiValentino2021s8, Mokeddem2025, Sabogal2025, Sabogal2025a, Nunes2020, BWRibeiro2026, Avila2025, Bom2026}. 
Moreover, with these best-fit values, we study the behavior of $f\sigma_8$ to understand if there is evidence for the suppression of cosmic structure growth at scales larger than 50 Mpc $h^{-1}$ in comparison with the $\Lambda$CDM prediction. Additionally, we study if the introduction of a scale-dependent term in the density contrast evolution equation can shed light on the solution to the intriguing challenges in modern cosmology, i.e., the $S_8$ and $H_0$ tensions~\citep{Sabogal2024, DiValentino2021,Perivolaropoulos2022,DiValentino2021s8,Nunes2021, Bargiacchi2023, Adil2023, Akarsu2025a, Colgain2025}.
Our approach is novel in that (i)~we consider a class of modified gravity models where the equation for the cosmic evolution of the matter fluctuations contains an explicit dependence on the scale $k$; (ii)~our statistical analyses include a set of uncorrelated $[f\sigma_8](z)$ measurements~\citep{Skara2019}; and (iii)~we include CMB data in our statistical analyses, combined with additional observables, namely CC, SN, and $f\sigma_8(z)$. 
In this context, the CMB data measured by the \textit{Planck satellite}~\citep{Planck2018} consists of the temperature and polarization anisotropy power spectra, together with their cross-spectra from the 2018 legacy release, 
as well as the reconstruction of CMB lensing. 

This work is organized as follows: In Section \ref{sec2}, we present the methodology and datasets used in this study. Our results are presented and discussed in Section \ref{sec3}, and our conclusions are presented in Section~\ref{sec4}.


\section{Parameterization of the scale-dependent 
growth of structures} \label{sec2}

The Linear Perturbation Theory for cosmological perturbations describes the evolution of matter density fluctuations $\delta_m(\textbf{r},a)$~(see, 
e.g.,~\cite{Coles1996, Avila2021, Skara2019, Marques20, Alonso2023, Shekhar2024, Franco2025b}), which is defined as 
\begin{equation}
\label{contrast}
\delta_m(\textbf{r},t) \equiv \frac{\rho_m(\textbf{r},t)- \Bar{\rho}_m(t)}{\Bar{\rho}_m(t)} \,,
\end{equation}
where $\rho_m(\textbf{r},t)$ is the matter density at position $\textbf{r}$ and cosmic time $t$, and $\Bar{\rho}_m(t)$ is the 
background matter density at the same epoch.

For sub-horizon scales and within the quasi-static approximation, the evolution of linear matter density perturbations is governed by
\begin{equation}
\label{eq:edo}
\ddot \delta_m(t) + 2 H(t)\,\dot \delta_m(t) 
- 4 \pi G\, \bar{\rho}_m(t)\,\delta_m(t) = 0 \,,
\end{equation}
where $H(t)\equiv \dot{a}(t)/a(t)$ is the Hubble parameter and $G$ is Newton's gravitational constant.

From the above equation, one can define the normalized growth rate of cosmic structures, $[f\sigma_8](a)$, defined as
\begin{equation}
    [f\sigma_8](a) \equiv \frac{\sigma_{8,0}}{\bar{\delta}_m(1)} \left[\frac{d \hspace{0.01cm} \delta_m(a)}{d \hspace{0.01cm} \ln(a)} \right],
\end{equation}
where $a$ is the scale factor and 
$\sigma_{8,0}$ is the variance of the matter fluctuations at the scale of 8 $h^{-1}$Mpc evaluated at $z=0$~\citep{Coles1996,Franco2025a}. 

Equation~(\ref{eq:edo}) is obtained assuming that the scale of the matter perturbations is significantly smaller than the Hubble scale.
However, this assumption is not valid on scales larger than 50 $h^{-1}$ Mpc, scales also interesting for studying the growth of structures.
Therefore, it is appropriate to investigate the introduction of a scale-dependent term in equation~(\ref{eq:edo}) to describe the time evolution of the linear matter density contrast (e.g., ~\citep{Dent2009, Perivolaropoulos2010, Denkiewicz2017, Denkiewicz2019, Sanchez2010})
\begin{equation}\label{edo_parametrization}
\ddot \delta_m(t) + 2 H(t)\,\dot \delta_m(t) - \frac{4 \pi\,G\, \bar{\rho}_m(t)\,\delta_m(t)}{1 + \xi(t,k)} = 0 \,,
\end{equation}
where we define
\begin{equation}\label{xi}
\xi(t,k) \equiv \frac{3\, H_0^2\,\Omega_{m,0}}{a(t)\,c^2\,k^2}\,.
\end{equation}
Solutions of equation~(\ref{edo_parametrization}) represent the scale-dependent evolution of the growth function as proposed by~\cite{Dent2009}.

The growth equation can be written in the standard form by introducing an effective gravitational coupling,
\begin{equation}
G_{\rm eff}(t,k) \equiv \frac{G}{1+\xi(t,k)}.
\end{equation}

Substituting this definition into equation~(\ref{edo_parametrization}), the evolution equation for the matter density contrast becomes
\begin{equation}
\ddot{\delta}_m
+2H\dot{\delta}_m
-4\pi G_{\rm eff}(t,k)\,\bar{\rho}_m\,\delta_m = 0 \,.
\end{equation}

It is also convenient to define the dimensionless modification to the gravitational coupling as
\begin{equation}
\mu(t,k)\equiv\frac{G_{\rm eff}(t,k)}{G}
=\frac{1}{1+\xi(t,k)}
=
\left(
1+\frac{3\, H_0^2\,\Omega_{m,0}}{a(t)\,c^2\,k^2}
\right)^{-1}.
\end{equation}
This parametrization makes explicit that the deviation from the standard gravitational coupling depends on both cosmic time and spatial scale.

In the small-scale limit, $k\rightarrow\infty$, one has $\xi\rightarrow0$, such that $G_{\rm eff}\rightarrow G$,
recovering the standard growth equation of General Relativity. Conversely, on sufficiently large scales, where $\xi>0$, the effective gravitational coupling satisfies
$G_{\rm eff}<G$, leading to a suppression of the gravitational source term and, consequently, to a slower growth of matter density perturbations. This form is particularly useful because it allows the model to be directly compared with phenomenological modified-gravity parameterizations in which the growth of structure is described through a time- and scale-dependent effective Newton's constant, $G_{\rm eff}(a,k)$.

Because the spacetime background of our scale-dependent model is determined by GR, the Hubble parameter is the same as that one given by the $\Lambda$CDM model (long after radiation domination), that is $H(a) = H_0 \sqrt{\Omega_{m,0}  \, a^{-3} + \Omega_{\Lambda, 0}}$, where $\Omega_{m,0}$ is the matter density parameter and 
$\Omega_{\Lambda, 0}$ is the dark energy density parameter, both evaluated at $z=0$. 
For a spatially flat universe, this parameter can be rewritten as $ \Omega_{\Lambda, 0} \equiv 1 - \Omega_{m,0}$. 

The confrontation of current observational data with the solutions of equation~\ref{edo_parametrization}) will assess the viability of this new scenario within GR, in which the growth of large-scale structures is scale-dependent while the background spacetime evolution remains consistent with GR.

To quantify the scale-dependent correction, we promote equation~(\ref{xi}) to 
$\xi(t,k)\rightarrow A \,\xi(t,k) \,$,
and, instead of sampling the amplitude $A$ directly, we introduce the transformed parameter\footnote{The factor $10^{4}$ is introduced as a numerical normalization to improve the efficiency of the parameter exploration. Since $\xi(t,k)$ is typically of order $10^{-5}$ on the linear scales considered. This normalization maps the region of interest into values of $B$ of order unity, leading to a more efficient MCMC sampling while leaving the underlying model unchanged.}
\begin{equation}\label{def-B}
B \equiv \ln\left(1+\frac{A}{10^{4}}\right) \,,
\end{equation}
equivalently,
\begin{equation}
A=10^{4}\left(e^{B}-1\right) \,.
\end{equation}
This transformation guarantees that the amplitude satisfies $A \geq 0$ while preserving the $\Lambda$CDM limit, which is recovered for $B = 0$. 
Moreover, it enables an efficient exploration of the parameter space by allowing the Markov chains to sample a logarithmic measure of the amplitude rather than the amplitude itself.

We implemented the theoretical model using the Boltzmann solver \texttt{CLASS}~\citep{Blas2011} and performed Markov Chain Monte Carlo (MCMC) analyses with the publicly available sampler \texttt{MontePython}~\citep{Audren2013, Brinckmann2019}, requiring a Gelman--Rubin convergence criterion~\citep{Gelman1992} of $R-1 \leq 10^{-2}$ for all chains.

We sampled the cosmological parameter set
\begin{equation}
\left\{
\omega_b,\,
\omega_{\rm cdm},\,
\tau_{\rm reio},\,
100\,\theta_s,\,
\ln(10^{10}A_s),\,
n_s,\,
B
\right\},
\end{equation}
where the first six correspond to the baseline $\Lambda$CDM cosmological parameters. These are the present-day physical baryon density ($\omega_b \equiv \Omega_b h^2$), the physical cold dark matter density ($\omega_{\rm cdm} \equiv \Omega_{\rm cdm} h^2$), the optical depth to reionization ($\tau_{\rm reio}$), the angular size of the sound horizon at recombination ($\theta_s$), the amplitude of the primordial scalar power spectrum ($A_s$), and the scalar spectral index ($n_s$). Flat priors were adopted over the ranges
$0.01 \leq \omega_b \leq 1.0,\qquad
0.01 \leq \omega_{\rm cdm} \leq 1.0,\qquad
0.004 \leq \tau_{\rm reio} \leq 0.8,
\,
0.2 \leq n_s \leq 2.0,\qquad
1.0 \leq \ln(10^{10}A_s) \leq 5.0,\qquad
0.5 \leq 100\,\theta_s \leq 2.0
$. 

For the new parameter $B$, we assume a flat prior 
\begin{equation}
0 \leq B \leq 3 \,,
\end{equation}
corresponding approximately to $0 \leq A\,\xi(t,k)\leq 2$.
Finally, we fix the comoving wavenumber to $k=0.1\,h\,\mathrm{Mpc}^{-1}$, since the effective gravitational coupling now depends explicitly on scale. This choice ensures that the analysis is performed within the linear regime, where linear perturbation theory is valid and has been extensively studied. 
Other reasonable choices of $k$ have no appreciable effect on our results.

Within this parametrization, the standard $\Lambda$CDM growth equation is exactly recovered for $B=0$, implying $A=0$, and therefore the absence of the scale-dependent correction. 
Conversely, finding a value $B \ne 0$ with high statistical significance would imply that a scale-dependent correction is required in the matter density contrast equation.

For the implementation of our analyses we first solve equation~(\ref{edo_parametrization}) to obtain the predicted $[f\sigma_8](z)$ function and then compare it against the $\Lambda$CDM\ baseline. The best-fit values for $H_0$, $\Omega_{m,0}$, $\sigma_{8,0}$, $A$, and the absolute magnitude\footnote{When SN data are included, we also impose $-30 \leq M_B \leq -10$.}, $M_B$, were inferred from a joint analysis combining CC, SN, $[f\sigma_8](z)$, and CMB datasets, described in detail in Section~\ref{datasets}. The MCMC chains were analyzed using the \texttt{GetDist} package\footnote{\url{https://github.com/cmbant/getdist}} to extract best-fit values, 1D posteriors, and 2D marginalized probability contours. 
The resulting $[f\sigma_8](z)$ function was subsequently used to investigate a possible suppression of large-scale cosmic structure growth.

\subsection{Data sets} \label{datasets}

In this work, we consider four datasets: 
\begin{itemize}
\item \textit{Redshift-Space Distortions} ($f\sigma_8$): We use a compilation of 35 uncorrelated measurements of the growth rate, $[f\sigma_8](z)$, as provided in~\citep{Skara2019}. These measurements are derived from redshift-space distortions in galaxy clustering and directly probe the growth of cosmic structures, making them a key observable for testing deviations from the standard $\Lambda$CDM growth scenario. 

\item \textit{Cosmic Chronometers} (\textbf{CC}): We include 31 measurements of the Hubble parameter, $H(z)$, obtained using the cosmic chronometer technique~\citep{Niu24}. This method relies on the differential age evolution of passively evolving galaxies to provide direct, model-independent estimates of the expansion rate, thereby offering robust constraints on the background cosmology. 
    
\item \textit{Type Ia Supernovae} (\textbf{SN}): SNe Ia distance modulus measurements (converted to luminosity distance data) from the PantheonPlus sample, which consists of 1550 supernovae spanning a redshift range from $0.01$ to $2.26$~\citep{Pantheon2022}.

\item \textit{Cosmic Microwave Background} (\textbf{CMB}): Temperature and polarization anisotropy measurements of the CMB power spectra (as well as their cross-spectra) from the Planck 2018 legacy data release (PR3). In particular, we use the high-$\ell$ \texttt{Plik} likelihood for TT ($30 \leq \ell \leq 2508$), TE, and EE ($30 \leq \ell \leq 1996$), as well as the low-$\ell$ TT-only ($2 \leq \ell \leq 29$) and EE-only ($2 \leq \ell \leq 29$) \texttt{SimAll} likelihoods~\citep{Planck:2019nip}. In addition, we also use the \texttt{Plik} CMB Planck lensing measurements~\citep{Planck:2018lbu}, reconstructed from the temperature 4-point correlation function.
\end{itemize}

\section{Main Results and Discussion} \label{sec3}

\begin{figure*}[!htbp]
\centering 
\hspace{-0.4cm}
\includegraphics[width=0.95\linewidth]{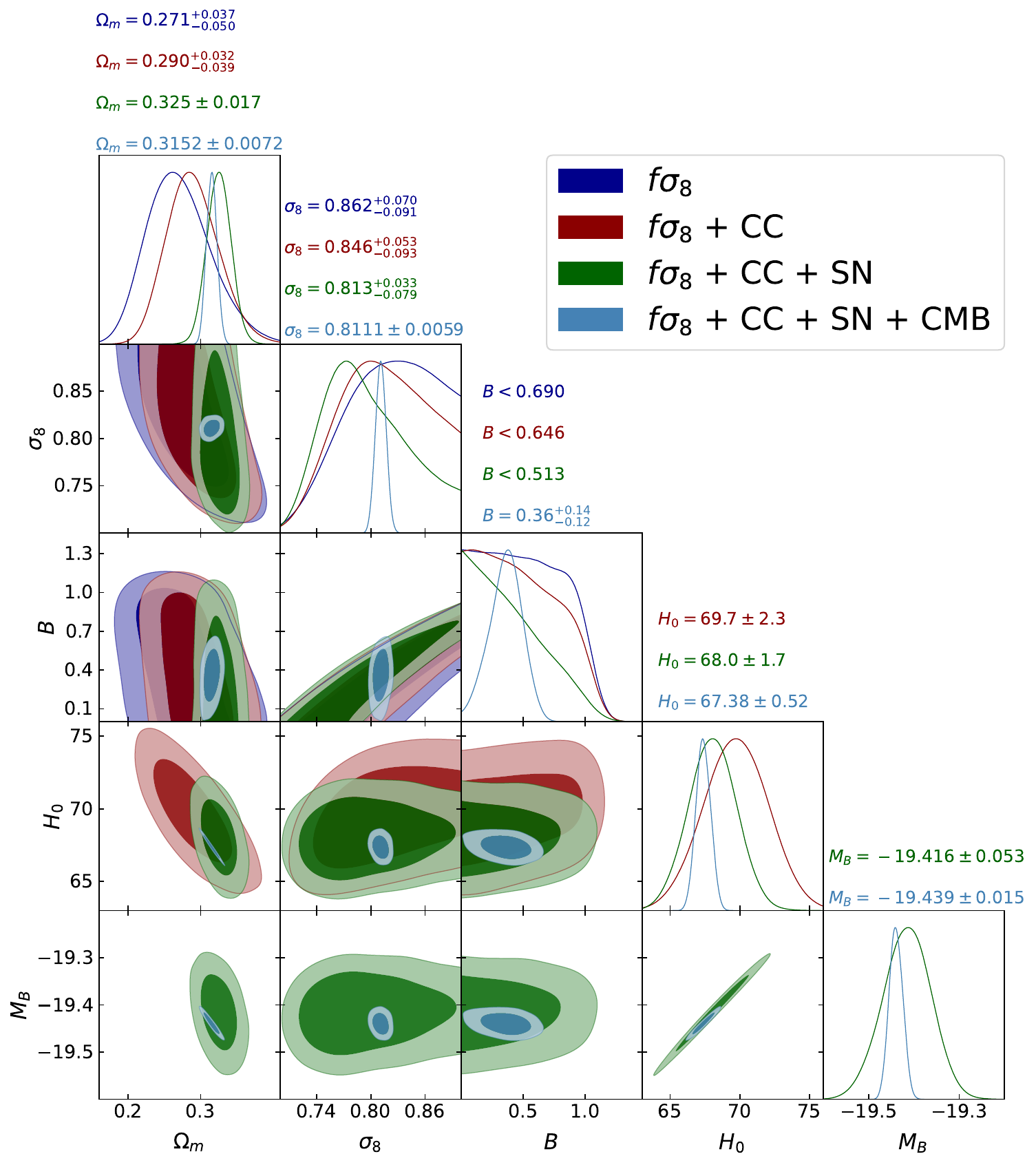}
\caption{Marginalized posterior distributions for a subset of the parameters investigated in this work. The dark blue curves correspond to the $f\sigma_8$ dataset, the red curves to $f\sigma_8$+CC, the green curves to $f\sigma_8$+CC+SN, and the light blue curves to the full $f\sigma_8$+CC+SN+CMB combination.}
\label{figure1}%
\end{figure*}

\begin{table*}[htpb]
\begin{centering}
\begin{tabular}{|c||c|c|c|} \hline
Parameters$\backslash$Datasets & $f\sigma_8$ & $f\sigma_8$+CC & $f\sigma_8$+CC+SN \\ \hline
$\Omega_{m,0}$ & $0.271^{+0.037}_{-0.050}$ & $0.290^{+0.032}_{-0.039}$   & $0.325 \pm 0.017$   \\ \hline
$\sigma_{8,0}$ &  $0.862^{+0.070}_{-0.091}$  &  $0.846^{+0.053}_{-0.093}$  & $0.813^{+0.033}_{-0.079}$  \\ \hline
$B$ & $< 0.690$ & $< 0.646$ & $< 0.513$  \\ \hline
$A$ & $< 9937.2$ & $< 9079.2$ & $< 6703.2$  \\ \hline
$H_0 \, [\mathrm{km/s/Mpc}]$ & - & $69.7 \pm 2.3$ & $68.0 \pm 1.7$  \\ \hline
$M_B$   & - & - & $-19.416 \pm 0.053$ \\ \hline

\end{tabular}
\caption{Marginalized parameter constraints, quoted as mean values with $68\%$ confidence intervals, for each dataset combination excluding CMB data.}
\label{table-results}
\end{centering}
\end{table*}

\begin{table}[htpb]
\begin{centering}
\begin{tabular}{|c||c|} \hline
Parameters$\backslash$Datasets  & $f\sigma_8$+CC+SN+CMB \\ \hline
$10^{2}\,\omega_{b}$ &  $2.24 \pm 0.014$  \\ \hline
$\omega_{cdm}$ & $0.12 \pm 0.001$  \\ \hline
$100\,\theta_s$ &  $1.04 \pm 0.0003$ \\ \hline
$\ln(10^{10} A_s)$ & $3.04 \pm 0.014$  \\ \hline
$\tau_{reio}$ &  $0.053 \pm 0.0072$ \\ \hline
$n_s$ & $0.965 \pm 0.003$  \\ \hline
$B$ &  $0.36^{+0.14}_{-0.12}$ \\ \hline
$A$ &  $4425 \pm 2000$ \\
\hline
\hline
$\Omega_{m,0}$ &  $0.3152 \pm 0.0072$  \\ \hline
$\sigma_{8,0}$ &  $0.8111 \pm 0.0059$ \\ \hline
$H_0 \, [\mathrm{km/s/Mpc}]$ &  $67.38 \pm 0.52$ \\ \hline
$M_B$   &  $-19.439 \pm 0.015$ \\ \hline
\end{tabular}
\caption{Marginalized constraints (mean values and $68\%$ confidence intervals) for the $f\sigma_8$+CC+SN+CMB dataset combination.}
\label{table-results2}
\end{centering}
\end{table}

Figure~\ref{figure1} displays the results from our MCMC analyses. The dark blue curves represent the analysis with only $[f\sigma_8](z)$ sample. The red curves correspond to the constraints obtained with $[f\sigma_8](z)$+CC data. The green curves represent the results with $[f\sigma_8](z)$+CC+SN data. The results presented in Table~\ref{table-results} show the impact of a scale-dependent modification on the growth of cosmic structures, as well as its consistency with current observational data.

A first notable aspect observed in Figure~\ref{figure1} is the progressive tightening and shift in the constraints on the matter density parameter, $\Omega_{m,0}$, as additional datasets are included. Using only $f\sigma_8$ data, we find $\Omega_{m,0} = 0.271^{+0.037}_{-0.050}$, which is relatively low and characterized by large uncertainties. This reflects the well-known degeneracy between $\Omega_{m,0}$ and $\sigma_{8,0}$ in growth-only measurements. 
The inclusion of CC slightly increases the central value and reduces uncertainties, while the addition of SN data leads to a significant tightening, yielding $\Omega_{m,0} = 0.325 \pm 0.017$. This value is in good agreement with standard $\Lambda$CDM constraints, indicating that background expansion data strongly anchor the matter density.

A complementary trend is observed for the amplitude of matter fluctuations, $\sigma_{8,0}$. When only $f\sigma_8$ data are considered, $\sigma_{8,0}$ assumes relatively high values, $\sim 0.86$, albeit with large uncertainties. As additional datasets are included, the central value shifts downward to $\sim 0.83$, with improved precision. This behavior is consistent with the breaking of degeneracies between $\sigma_{8,0}$ and $\Omega_{m,0}$, and may be interpreted as a mild alleviation of the well-known $\sigma_8$ tension, although uncertainties remain sufficiently large to prevent a definitive conclusion.

The most relevant parameter for this analysis is the scale amplitude $A$, which quantifies deviations from the standard scale-independent growth predicted by $\Lambda$CDM. In the following dataset combinations we obtain only upper limits on $A$: $A < 9000$ (95\% CL) for $f\sigma_8$ and $f\sigma_8$+CC, and $A < 6703.2$ when SN data are included. 
The absence of a statistically significant detection of a non-zero $A$ indicates that these data do not favor scale-dependent growth within the adopted parametrization. 

From a physical perspective, this result implies that any scale dependence in the growth of structures -at least of the form encoded in equation~(\ref{edo_parametrization})- must be subdominant within the range of scales probed by the data. Since the parametrization effectively suppresses the gravitational clustering term via the factor $(1+\xi)^{-1}$, the constraint on $A$ can be interpreted as an upper bound on possible large-scale suppression effects. In other words, the growth of cosmic structures remains consistent with the standard General Relativity prediction to within $\mathcal{O}(10\%)$ corrections in the effective clustering strength.

The inferred values of the Hubble constant, $H_0 = 69.7 \pm 2.3$ km/s/Mpc (CC) and $H_0 = 68.0 \pm 1.7$ km/s/Mpc (CC+SN), are consistent with late-universe measurements and lie below local distance ladder estimates. 
The inclusion of SN data shifts $H_0$ slightly downward and improves precision, reflecting the constraining power of geometric probes. 
Importantly, the introduction of the scale-dependent parameter $A$ does not lead to a significant shift in $H_0$, suggesting that this extension does not alleviate the Hubble tension. The absolute magnitude parameter $M_B = -19.416 \pm 0.053$ is consistent with standard SN calibrations, indicating internal consistency of the combined dataset.

Overall, in light of the samples in study, these results suggest that a scale-dependent modification of the growth of structures, within the adopted phenomenological framework, is not required by current data. 
The standard $\Lambda$CDM scenario, characterized by scale-independent linear growth, remains a robust description of observations. 
Nevertheless, the obtained upper bounds on $B$ --or equivalently on $A$, see equation~(\ref{def-B})-- provide a quantitative limit on possible deviations and may serve as a benchmark for future high-precision surveys, which will probe larger volumes and wider ranges of scales, potentially revealing subtle scale-dependent effects beyond the reach of current datasets.

\textit{Impact of CMB Data on Scale-Dependent Growth Constraints}. 
The inclusion of CMB data leads to a substantial improvement in the precision of all cosmological parameters, as expected from the constraining power of early-universe observations. In particular, the baseline $\Lambda$CDM parameters are now tightly constrained and fully consistent with previous analyses based on CMB data alone.

According to our results, summarized in Table~\ref{table-results2} and Figure~\ref{figure1}, 
the baryon density, $10^{2}\,\omega_b = 2.24 \pm 0.014$, and cold dark matter density, $\omega_{\rm cdm} = 0.12 \pm 0.001$, are both in excellent agreement with standard Planck results, indicating that the introduction of the scale-dependent parameter $B$ does not significantly perturb early-universe physics. Similarly, the angular acoustic scale, $100\,\theta_s$, is determined with sub-percent precision, preserving the standard distance to last scattering.

The primordial parameters, $\ln(10^{10}A_s) = 3.04 \pm 0.014$ and $n_s = 0.965 \pm 0.003$, also remain fully consistent with a nearly scale-invariant primordial spectrum, while the optical depth $\tau_{\rm reio} = 0.053 \pm 0.0072$ matches recent low-$\tau$ determinations. These results confirm that the proposed scale-dependent growth parametrization does not introduce any significant tension with early-universe observables.

At the background and late-time level, we obtain $\Omega_{m,0} = 0.3152 \pm 0.0072$ and $H_0 = 67.38 \pm 0.52$ km/s/Mpc, both in remarkable agreement with CMB-driven $\Lambda$CDM constraints. The value of $\sigma_{8,0} = 0.8111 \pm 0.0059$ is also tightly constrained and slightly lower than the values obtained without CMB data, reflecting the strong anchoring effect of early-universe physics on the amplitude of matter fluctuations. This value remains compatible with large-scale structure measurements and may indicate a mild preference toward reduced clustering amplitude, although not at a statistically significant level.

The most striking result concerns the scale amplitude parameter $A = 4425 \pm 2000$, or equivalently $B=0.36^{+0.14}_{-0.12}$. 
Unlike the previous dataset combinations, which only provided upper bounds, the inclusion of CMB data leads to a non-zero central value with a $2.2 \,\sigma$ indication for $A > 0$. This result suggests a mild preference for a scale-dependent suppression of the growth of cosmic structures.

From a physical standpoint, a positive value of $A$ enhances the quantity $\xi(t,k)$, effectively reducing the strength of gravitational clustering through the factor $(1+\xi)^{-1}$ in equation~(\ref{edo_parametrization}). 
This translates into a suppression of the growth rate at large scales (small $k$), while leaving small-scale growth essentially unaffected. Therefore, our results may be hinting at a scale-dependent weakening of structure formation, which could be interpreted as an effective large-scale modification of gravity or as a phenomenological description of relativistic corrections beyond the Newtonian approximation.
Importantly, this feature emerges only when combining early- and late-universe probes, highlighting the role of CMB data in breaking degeneracies present in growth-only analyses. In particular, the tight constraints on $\Omega_{m,0}$ and $\sigma_{8,0}$ provided by the CMB data reduce the allowed parameter space, making it possible to detect subtle deviations encoded in $A$. Despite this intriguing hint, the statistical significance remains modest, and the result should be interpreted with caution. Systematic uncertainties in large-scale structure data, as well as potential modeling assumptions, may still play a role. Nevertheless, if confirmed by more precise data from future surveys, such a scale-dependent suppression could have profound implications, potentially pointing toward beyond-$\Lambda$CDM physics, including modified gravity scenarios or large-scale relativistic effects.

\begin{figure}[!htbp]
\centering
\hspace{-0.51cm}
\includegraphics[width=1.05\linewidth]{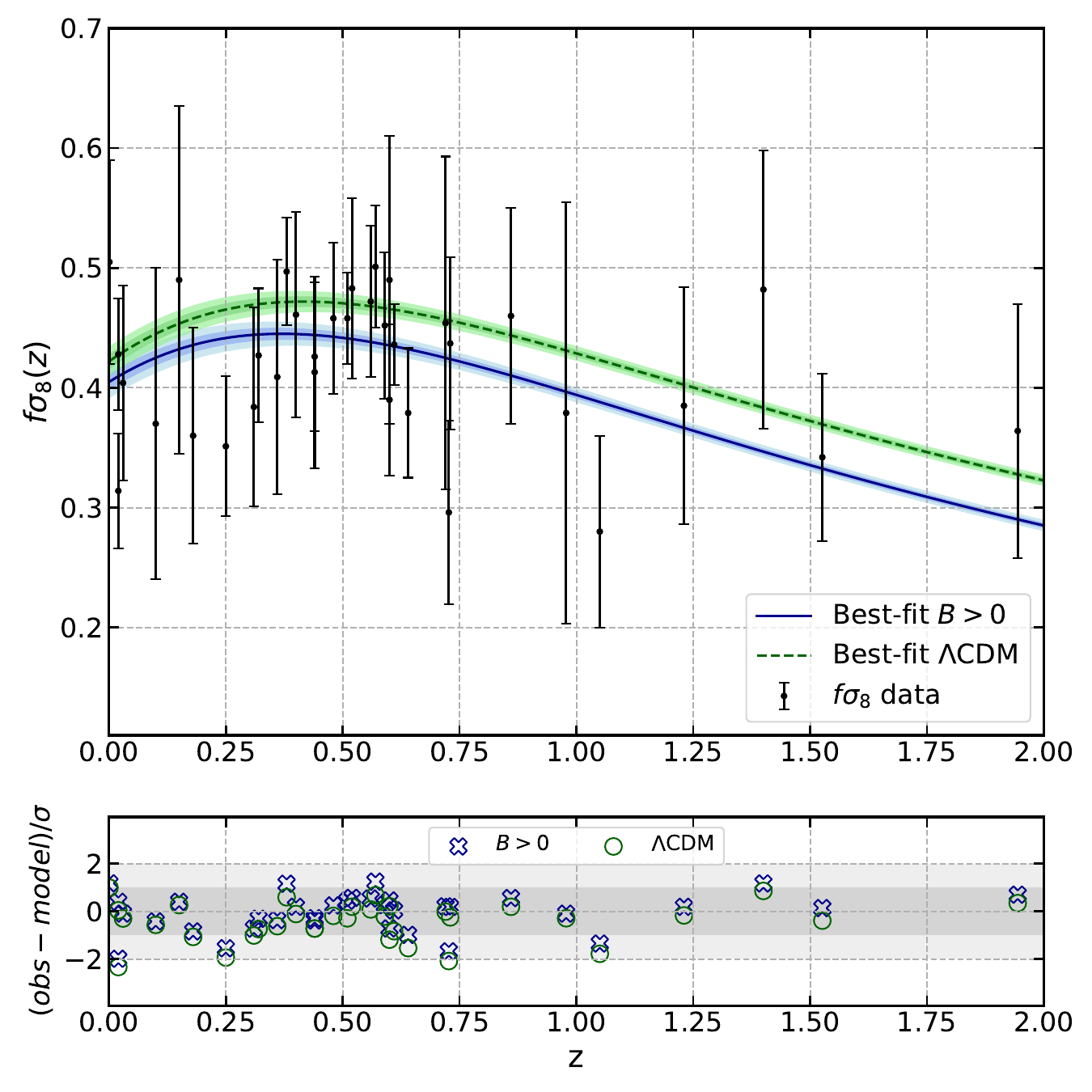}
\caption{Comparison between $[f\sigma_{8}](z)$ prediction using $f\sigma_{8}$+CC+SN+CMB data for the scale-dependent model and the $\Lambda$CDM model. The shaded areas represent the 
$2\,\sigma$ CL region. The bottom panel shows the relative difference between the data points and the two models analyzed.}
\label{figure2}
\end{figure}

In Figure~\ref{figure2}, we compare the $[f\sigma_8](z)$ predictions of the $\Lambda$CDM model (green) with those of the scale-dependent model for $B = 0.36$ (blue). 
The shaded regions indicate the $2\,\sigma$ confidence intervals (i.e., $95\%$ CL), while the black data points with error bars correspond to the compilation of $[f\sigma_8](z)$ measurements from~\citep{Skara2019}.

The lower panel of Figure~\ref{figure2} shows the relative difference between the observational data and the theoretical predictions of both models. At low redshift, $z \lesssim 0.5$, the relative differences remain close to zero, indicating that both models provide an adequate description of the data in this regime. At higher redshifts, $z \gtrsim 0.5$, deviations become more pronounced. In particular, the $\Lambda$CDM model shows a reduced ability to accurately reproduce the $[f\sigma_8](z)$ measurements, while the scale-dependent model exhibits a better overall agreement. This behavior highlights the impact of the scale-dependent correction, which effectively modifies the growth of structures and leads to observable differences at intermediate and high redshifts.

In general, the scale-dependent model fits the observational data slightly better across the entire redshift range, primarily driven by the $f\sigma_8$ and \texttt{Planck\_highl\_TTTEEE} datasets, as shown in Table \ref{table-deltachi2}. The $f\sigma_8$ measurements yield $\Delta \chi^2_{\rm min} = -6.42$, reflecting an improved capacity to describe the growth rate of structure formation, while the high-multipole Planck temperature and polarization data contribute with an additional $\Delta \chi^2_{\rm min} = -0.28$. The SN dataset demonstrates a modest improvement with $\Delta \chi^2_{\rm min} = -1.16$, whereas the CC constraint shows fit equivalence with $\Delta \chi^2_{\rm min} = +0.21$. Among the Planck low-multipole measurements, the EE polarization channel favors $\Lambda\mathrm{CDM}$ with $\Delta \chi^2_{\rm min} = +0.05$, while the low-$\ell$ TT component and lensing reconstruction display differences with $\Delta \chi^2_{\rm min} = +1.02$. Collectively, these improvements yield a total joint constraint of $\Delta \chi^2_{\rm min} = -6.04$, which highlights that the scale-dependent model provides better fits than $\Lambda\mathrm{CDM}$ across the entire range of redshifts.

However, the Bayesian evidence calculation\footnote{To compute the Bayes factors, we employ the \texttt{MCEvidence} package~\citep{Heavens:2017hkr,Heavens:2017afc}, which is publicly available at the following link: \url{https://github.com/yabebalFantaye/MCEvidence}.} tells a more balanced situation. 
According to the Kass \& Raftery scale~\citep{Kass:1995loi}, the model comparison yields an inconclusive result, $|\ln \mathcal{B}_{ij}|<1$. Although the scale-dependent model achieves a 
cumulative~\footnote{that is, the value obtained considering all the datasets} $\chi^2_{\rm min}$ value lower than in 
$\Lambda\mathrm{CDM}$ model, the Bayesian framework penalizes it for introducing an additional free parameter and also considers how efficiently the parameter space is explored. Therefore, with a value of $\ln \mathcal{B}_{ij} = -0.87$, our analysis suggests that the scale-dependent model is equally viable as $\Lambda\mathrm{CDM}$.

\begin{table}[htpb]
\begin{centering}
\begin{tabular}{|c|c|} \hline
Dataset & $\Delta \chi^2_{\rm min}$ \\ \hline
\hline
$f\sigma_8$ & $-6.42$ \\ \hline
CC & $+0.21$ \\ \hline
SN & $-1.16$ \\ \hline
\hline
\texttt{Planck\_highl\_TTTEEE} & $-0.28$ \\ \hline
\texttt{Planck\_lowl\_EE} & $+0.05$ \\ \hline
\texttt{Planck\_lowl\_TT} & $+1.02$ \\ \hline
\texttt{Planck\_lensing} & $+0.55$ \\ \hline
\hline
Total & $-6.04$ \\ \hline
$\ln \mathcal B_{ij}$ & $-0.87$\\ \hline
\end{tabular}
\caption{$\Delta \chi^2_{\rm min} = \chi^2_{{\rm min} \,(\text{This work})} - \chi^2_{{\rm min}\,(\Lambda\mathrm{CDM})}$ for each dataset and the total joint analysis.}
\label{table-deltachi2}
\end{centering}
\end{table}

More specifically, the comparison at higher redshifts, $z \gtrsim 0.5$, between the predicted values of $f\sigma_8$ in both models reveals a more pronounced suppression of the growth rate of large-scale structures in the scale-dependent scenario. This behavior reflects the fact that the scale-dependent correction becomes increasingly relevant at earlier times (larger $z$), when the factor $\xi(t,k) \propto (1+z)/k^2$ is enhanced. As a consequence, the effective gravitational clustering is reduced, leading to a slower growth of matter perturbations.

This result implies that, during the matter-dominated epoch, the suppression of structure growth is stronger than in the late-time, dark-energy-dominated regime. Physically, this is expected within this parametrization, since the scale-dependent term decreases as the universe expands, gradually restoring the standard $\Lambda$CDM behavior at low redshifts. Such a suppression of the growth rate has been widely discussed in the literature in connection with current cosmological tensions, particularly those involving the amplitude of matter fluctuations~\citep{Nguyen2023, Ruiz2015}. In this context, introducing a scale-dependent modification to the growth of structures provides a natural mechanism to reduce the clustering amplitude at large scales, and may therefore offer a phenomenological route toward alleviating the $S_8$ tension.

To further quantify this aspect, we estimate the $S_8$ parameter within the present framework. Recall that $S_8$ is defined as
\begin{equation}
\label{s8}
S_8 \equiv \sigma_{8,0} \left( \frac{\Omega_{m,0}}{0.3} \right)^{1/2} \,.
\end{equation}
From our MCMC analysis, we obtain $S_8 = 0.831 \pm 0.011$, which is in excellent agreement with the value reported by the Planck collaboration, $S_8 = 0.832 \pm 0.013$~\citep{Planck2018}. Similarly, the inferred Hubble constant, $H_0 = 67.38 \pm 0.52 \,\mathrm{km\,s^{-1}\,Mpc^{-1}}$, is fully consistent with the Planck result, $H_0 = 67.36 \pm 0.54 \,\mathrm{km\,s^{-1}\,Mpc^{-1}}$~\citep{Planck2018}. These agreements indicate that the proposed scale-dependent extension preserves the successful description of early- and late-time background observables.

The most significant result of this analysis concerns the constraint on the scale amplitude parameter $A$. Using the full dataset, we find $A = 4425 \pm 2000$, or equivalently $B = 0.36^{+0.14}_{-0.12}$, corresponding to a $2.2 \,\sigma$ preference for a non-zero value. This constitutes a mild but intriguing indication that the growth of cosmic structures may deviate from strict scale independence. 
In physical terms, a positive value of $A$ enhances the role of the correction term $\xi(t,k)$, effectively weakening gravitational clustering on large scales and leading to the observed suppression in $f\sigma_8(z)$. 
It is important to emphasize, however, that the statistical significance of this result remains limited, and it should therefore be interpreted with caution. Residual systematics in large-scale structure data, as well as assumptions in the modeling, may still influence the inferred value of $A$. Nevertheless, this finding is consistent with previous studies reporting suppressed growth rate of cosmic structures, and may point toward new physics beyond the standard $\Lambda$CDM paradigm~\citep{Ruiz2015, Nguyen2023}.

\section{Final Remarks} 
\label{sec4}

The growth of cosmic structures, driven by gravitational instability acting on primordial density perturbations, provides one of the most powerful probes for testing the standard $\Lambda$CDM cosmology and exploring possible deviations from General Relativity and alternative models of dark energy.
In this work, we investigate the possibility of a scale-dependent suppression of the growth of large-scale structures by introducing a phenomenological modification to the evolution equation of the linear matter density contrast. While the standard scale-independent description provides an excellent approximation on sub-horizon scales, relativistic corrections become increasingly relevant toward horizon-sized scales~\citep{Dent2009}. Motivated by this behavior, we extend the linear growth equation by introducing a scale-dependent function, $\xi(t,k)$, which modifies the effective gravitational coupling governing the growth of matter perturbations.

We constrained this model using Markov Chain Monte Carlo (MCMC) analyses combining $f\sigma_8(z)$, cosmic chronometers $H(z)$, supernovae luminosity distances, and CMB data. 
The parameter $B$ controls the amplitude of the scale-dependent correction, with $B=0$ recovering the standard $\Lambda$CDM growth evolution. 
Our results, summarized in Table~\ref{table-results2} and Figures~\ref{figure1} and~\ref{figure2}, show that the scale-dependent model leads to a suppression of the growth rate relative to $\Lambda$CDM, particularly at intermediate and high redshifts ($z \gtrsim 0.5$). 
This behavior arises naturally from the redshift and scale dependence of the correction term, which effectively weakens gravitational clustering at earlier times and on large scales.
In fact, using the full dataset, we obtain $B = 0.36^{+0.14}_{-0.12}$, which leads us to $A = 4425 \pm 2000$, corresponding to a $2.2 \,\sigma$ preference for a non-zero scale-dependent contribution. 
At the background level, we find $S_8 = 0.831 \pm 0.011$ and $H_0 = 67.38 \pm 0.52\,\mathrm{km\,s^{-1}\,Mpc^{-1}}$, both in excellent agreement with Planck results~\citep{Planck2018}. 
This demonstrates that the proposed extension preserves the successful description of the background expansion and early-universe observables. 
At the same time, it also indicates that the model does not significantly alleviate the current $S_8$ or $H_0$ tensions.

In summary, while the $\Lambda$CDM model remains fully consistent with the data, our analysis provides a mild indication that scale-dependent effects may play a role in the growth of cosmic structures at very large scales. 
Future high-precision surveys \citep{Amendola2016, Zhan2018}, probing larger volumes and wider ranges of scales, will be essential to confirm the presence of such effects and to clarify their physical origin.

\section*{Acknowledgements}
FO thanks CAPES for the fellowship. MAS acknowledges support from CAPES and expresses gratitude to the Observatório Nacional for their hospitality during the development of this work. MAS also acknowledges support during the final stages of this work from the University of Trento and the Provincia Autonoma di Trento (PAT, Autonomous Province of Trento). 
 RCN thanks the financial support from the Conselho Nacional de Desenvolvimento Científico e Tecnológico (CNPq, National Council for Scientific and Technological Development) under the project No. 304306/2022-3, and the Fundação de Amparo à Pesquisa do Estado do RS (FAPERGS, Research Support Foundation of the State of RS) for partial financial support under the project No. 23/2551-0000848-3. FA thanks to Fundação de Amparo à Pesquisa do Estado do Rio de Janeiro (FAPERJ), Processo SEI-260003/001221/2025,  for the financial support. 
 AB acknowledges a CNPq fellowship.

\appendix


\bibliographystyle{elsarticle-harv} 
\bibliography{example}






\end{document}